\documentclass[11pt]{article}
\usepackage{moriond,epsfig}

\def\be{\begin{equation}}
\def\ee{\end{equation}}
\def\bea{\begin{eqnarray}}
\def\eea{\end{eqnarray}}

\newcommand{\gsim}{\mbox{ \raisebox{-1.0ex}{$\stackrel{\textstyle >}
{\textstyle \sim}$ }}}
\newcommand{\lsim}{\mbox{ \raisebox{-1.0ex}{$\stackrel{\textstyle <}
{\textstyle \sim}$ }}}

\begin{document}
\vspace*{4cm}
\title{A TeV scale model for neutrino mass, dark matter and baryon asymmetry}

\author{Mayumi Aoki$^1$, Shinya Kanemura$^{2}$ \footnote{Speaker},
  Osamu Seto$^{3}$ \footnote{Address after April 2009:
    Department of Architecture and Building Engineering,
    Hokkai-Gakuen University, Sapporo 062-8605, Japan}}

\address{$^1$Department of Physics,~Tohoku University,~Aramaki, Aoba, Sendai,~Miyagi 980-8578, Japan\\
$^2$Department of Physics, University of Toyama,3190 Gofuku, Toyama 930-8555, Japan\\
$^3$William I. Fine Theoretical Physics Institute, University of Minnesota,
     Minneapolis, MN 55455, USA}

\maketitle\abstracts{
  We discuss a TeV scale model which would explain neutrino
  oscillation, dark matter, and baryon asymmetry of the Universe
  simultaneously by the dynamics of the extended Higgs sector and TeV-scale
 right-handed neutrinos with imposed an exact $Z_2$ symmetry. 
 Tiny neutrino masses are generated at the three loop level, 
 a singlet scalar field is a candidate of dark matter, 
 and a strong first order phase transition is realized for successful
 electroweak baryogenesis.
 The model provides various discriminative predictions,
 so that it is testable at the current and future experiments.}

\section{Introduction}

Today, we know that a new model beyond the standard model (SM) must be considered
to understand the phenomena such as tiny neutrino masses and their mixing, 
the nature of dark matter (DM)
and baryon asymmetry of the
Universe.

In this talk, we discuss a model which would explain these problems simultaneously 
by an extended Higgs sector with TeV-scale right-handed (RH) neutrinos~\cite{aks}. 
Tiny neutrino masses are generated at the three loop level due to an
exact discrete symmetry, by which tree-level Yukawa couplings of neutrinos are prohibited.
The lightest neutral odd state under the discrete symmetry is a candidate of DM.  
Baryon number can also be generated at the electroweak phase transition
(EWPT) by additional CP violating phases in the Higgs sector~\cite{ewbg-thdm}.
In this framework, a successful model can be made without contradiction
of the current data.

Original idea of generating tiny neutrino masses via the radiative effect 
has been proposed by Zee~\cite{zee}. 
The extension with a TeV-scale  RH neutrino has been discussed in Ref.~\cite{knt},
where neutrino masses are generated at the three-loop level due to the exact $Z_2$
parity, and the $Z_2$-odd RH neutrino is a candidate of DM. This 
has been extended with two RH neutrinos to describe the neutrino data~\cite{kingman-seto}. 
Several models with adding baryogenesis have been considered in Ref.~\cite{ma}.
The following advantages would be in the present model:
(a)~all mass  scales are at most at the TeV scale without large hierarchy, 
(b)~physics for generating neutrino masses is connected with that for
DM and baryogenesis, 
(c)~the model parameters are strongly constrained by the current data, so
    that the model provides testable and discriminative prediction at future experiments.

\section{Model}

We introduce two scalar isospin doublets with hypercharge $1/2$ ($\Phi_1$ and $\Phi_2$),  
charged singlet fields ($S^\pm$), a real scalar singlet ($\eta$) and two
generation isospin-singlet RH neutrinos ($N_R^\alpha$ with $\alpha=1, 2$).
We impose an exact $Z_2$ symmetry to generate tiny neutrino masses
at the three-loop level, which we refer as $Z_2$. 
We assign $Z_2$-odd charge to $S^\pm$, $\eta$ and $N_R^\alpha$, while 
ordinary gauge fields, quarks and leptons and Higgs doublets are  $Z_2$ even.
In order to avoid the flavor changing neutral current, we impose
another (softly-broken) discrete symmetry ($\tilde{Z}_2$).
We assign $\tilde{Z}_2$ charges such that only $\Phi_1$ couples
to leptons whereas $\Phi_2$ does to quarks, as summarized in Table~\ref{discrete}.
The Yukawa coupling in our model~\cite{barger,type-X,logan}, which we refer to as the type-X~\cite{type-X},
is different from that in the minimal supersymmetric SM (MSSM).
\begin{table}[t]
\begin{center}
\centerline{
  \begin{tabular}{c|ccccc|cc|ccc}
   \hline
   & $Q^i$ & $u_R^{i}$ & $d_R^{i}$ & $L^i$ & $e_R^i$ & $\Phi_1$ & $\Phi_2$ & $S^\pm$ &
    $\eta$ & $N_{R}^{\alpha}$ \\\hline
$Z_2\frac{}{}$                ({\rm exact}) & $+$ & $+$ & $+$ & $+$ & $+$ & $+$ & $+$ & $-$ & $-$ & $-$ \\ \hline  
$\tilde{Z}_2\frac{}{}$ ({\rm softly\hspace{1mm}broken})& $+$ & $-$ & $-$ & $+$ &
                       $+$ & $+$ & $-$ & $+$ & $-$ & $+$ \\\hline
   \end{tabular}
 }
  \caption{Particle properties under the discrete symmetries.
 }
  \label{discrete}
\end{center}
\end{table}

As $Z_2$ is exact, the even and odd fields cannot mix.
Mass matrices for the $Z_2$ even scalars are diagonalized as in the
usual THDM by the mixing angles $\alpha$ and $\beta$, where $\alpha$
diagonalizes the CP-even states, and $\tan\beta=\langle \Phi_2^0
\rangle/\langle \Phi_1^0 \rangle$. 
The $Z_2$ even physical states are two CP-even ($h$ and $H$),
a CP-odd ($A$) and charged ($H^\pm$) states.
We here define $h$ and $H$ such that $h$ is always
the SM-like Higgs boson when $\sin(\beta-\alpha)=1$. 

\section{Neutrino Mass, Dark Matter, 1st Order Phase Transition}

\begin{figure}
  \begin{center}
     \includegraphics[width=.7\textwidth]{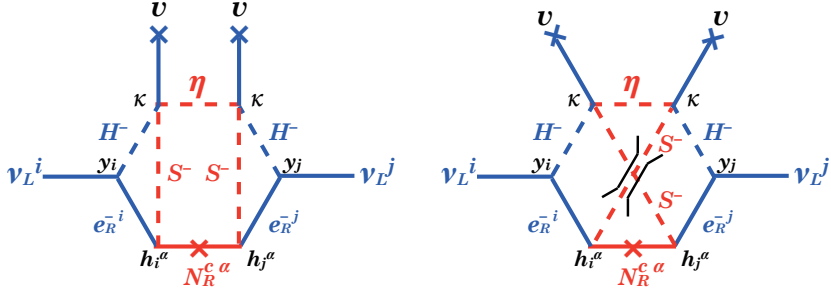}
  \caption{The diagrams for generating tiny neutrino masses.}
  \label{diag-numass}
    \end{center}
  \end{figure}
The LH neutrino mass matrix $M_{ij}$ is generated by the three-loop diagrams in Fig.~\ref{diag-numass}. 
To reproduce the neutrino data under the {\it natural} requirement on the coupling constant 
$h_e^\alpha \sim {\cal O}(1)$ in Fig.~\ref{diag-numass} and the  $\mu\to e\gamma$ results~\cite{lfv-data}, 
we find that $m_{N_R^\alpha}^{} \sim {\cal O}(1)$ TeV, 
$m_{H^\pm}^{} \lsim {\cal O}(100)$ GeV, $\kappa \tan\beta \gsim {\cal
O}(10)$, and $m_{S^\pm}^{}$ being several times 100 GeV. 
On the other hand, the LEP direct search results indicate 
$m_{H^\pm}^{}$ (and $m_{S^\pm}^{}$)  $\gsim 100$ GeV~\cite{lep-data}.  
In addition, with the LEP precision data for the $\rho$ parameter,  
the preferred values turn out to be  
$m_{H^\pm}^{} \simeq m_{H}^{}$ (or $m_{A}^{}$) $\simeq 100$ GeV
for $\sin(\beta-\alpha) \simeq 1$. 
Thanks to the Type-X Yukawa coupling~\cite{type-X}, such
a light $H^\pm$ is not excluded by the $b \to s \gamma$ data~\cite{bsgamma}.
Since we cannot avoid to include the hierarchy among $y_i^{\rm SM}$,  
we only require $h_i^\alpha y_i \sim {\cal O}(y_e) \sim 10^{-5}$ 
for values of $h_i^\alpha$. 
%
%
%

\begin{figure}
 \includegraphics[width=.5\textwidth]{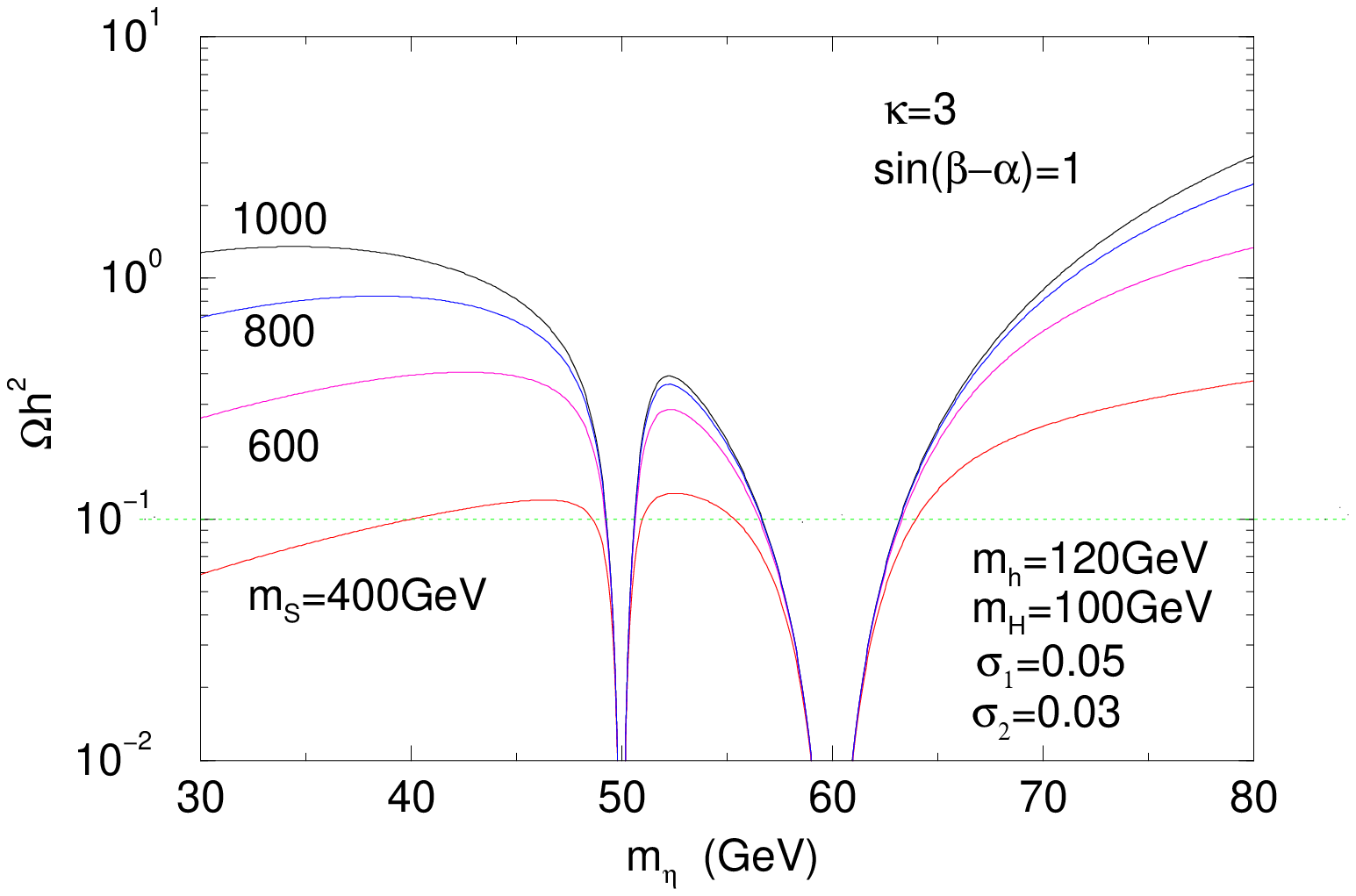}
 \includegraphics[width=.42\textwidth]{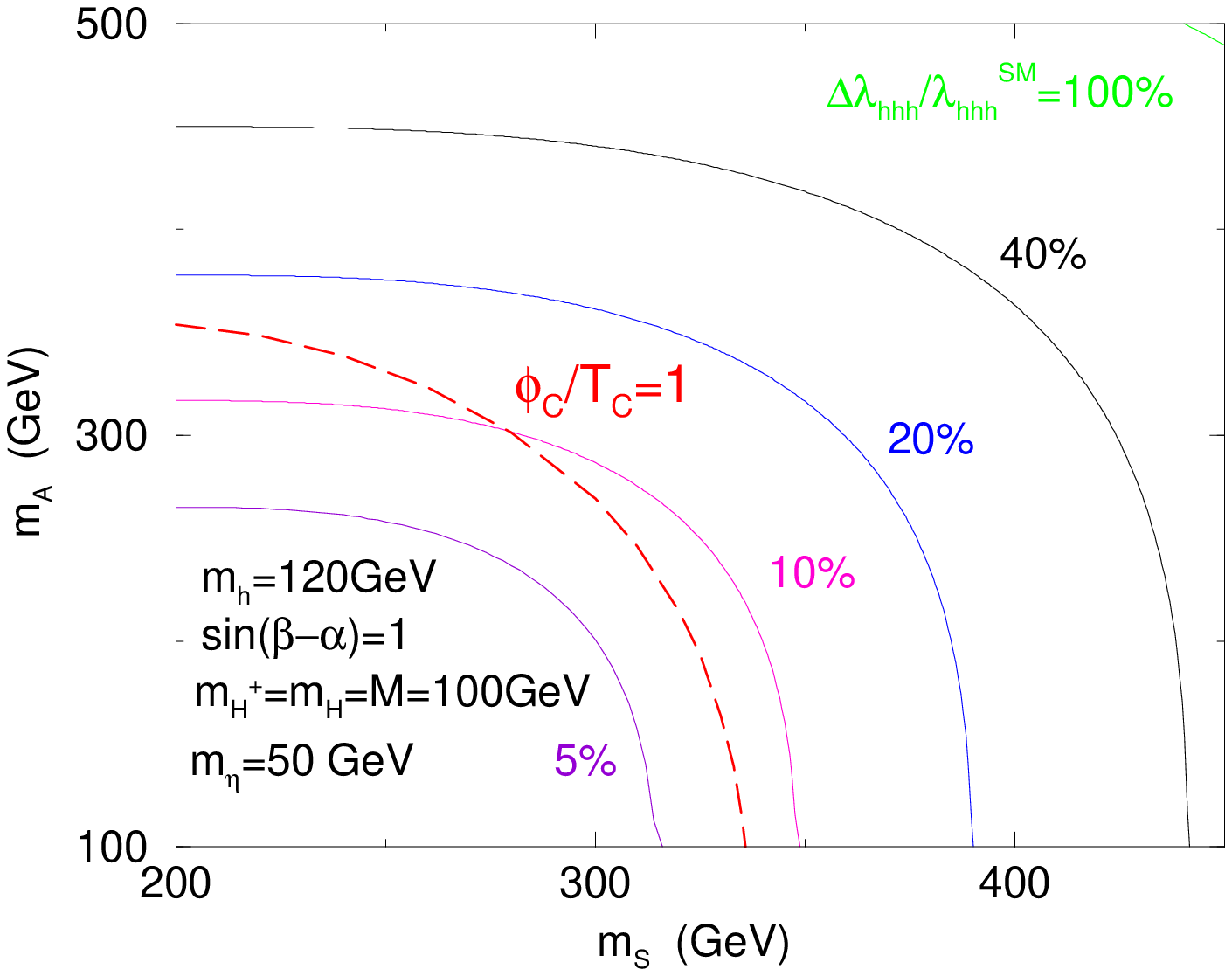}
  \caption{[Left figure] The relic abundance of $\eta$.
 [Right figure] The region of strong first order EWPT.
 Deviations from the SM value in the $hhh$ coupling
 are also shown. }
  \label{etaOmega}
\end{figure}

\indent
The lightest $Z_2$-odd particle is 
stable and can be a candidate of DM if it is neutral.
In our model, $N_R^\alpha$ must be heavy, so that 
the DM candidate is identified as $\eta$.
When $\eta$ is lighter than the W boson, $\eta$ dominantly annihilates 
into $b \bar{b}$ and $\tau^+\tau^-$ via tree-level $s$-channel
Higgs ($h$ and $H$) exchange diagrams, and into $\gamma\gamma$ via
one-loop diagrams.
From their summed thermal averaged annihilation rate $\langle \sigma v \rangle$,
the relic mass density  $\Omega_\eta h^2$ is 
evaluated.
Fig.~\ref{etaOmega}(Left) shows 
$\Omega_{\eta}h^2$ as a function of $m_\eta$. 
The data ($\Omega_{\rm DM} h^2 \simeq 0.11$) indicate that $m_\eta$ is around 40-65 GeV. %


The model satisfies the necessary 
conditions for baryogenesis.    
Especially, departure from thermal equilibrium can be 
realized by the strong first order EWPT.
For sufficient sphaleron decoupling in the broken phase,
it is required that~\cite{sph-cond} 
$\varphi_c/T_{c}  \gsim 1, \label{sph2}$
where $\varphi_c$ ($\neq 0$) and $T_c$ are the critical values of
$\varphi$ and $T$ at the EWPT.
In Fig.~\ref{etaOmega}(Right), the allowed region under this condition 
is shown. The condition is satisfied when
$m_{S^{\pm}}^{} \gsim 350$ GeV
for $m_A^{} \gsim 100$ GeV, 
$m_h \simeq 120$~GeV, $m_H^{} \simeq m_{H^\pm}^{} (\simeq 
M) \simeq 100$ GeV and $\sin(\beta-\alpha)\simeq 1$.

\section{Phenomenology}

A scenario which can simultaneously solve the three issues 
under the data~\cite{lfv-data,lep-data,bsgamma} would be 
\begin{eqnarray}
 \begin{array}{llll}
 \sin(\beta-\alpha) \simeq 1, &\!\!
 \kappa \tan\beta \simeq 30, &\!\!
 m_h = 120 {\rm GeV},     &\!\!
 m_H^{} \simeq m_{H^\pm} \simeq {\cal O}(100) {\rm GeV},    \\
  m_A \gsim {\cal O}(100) {\rm GeV},
   &\!\! m_{S^\pm}^{}\sim 400{\rm GeV},&\!\!
  m_{\eta} \lsim m_W^{},
  &\!\! m_{N_R^{1}} \simeq m_{N_R^{2}} \simeq 3 {\rm TeV}.\\
  \end{array} \nonumber
\end{eqnarray}
This is realized without assuming unnatural hierarchy among the
couplings. All the masses are  between ${\cal O}(100)$ GeV and ${\cal O}(1)$ TeV.
The discriminative phenomenological properties of this scenario are
discussed in details in Refs.~\cite{aks} and \cite{type-X}. We shortly summarize them in the following.

      The SM-like Higgs boson $h$ decays into $\eta\eta$ when $m_\eta < m_h/2$.
      The branching ratio is about 30\% for $m_\eta \simeq 43$ GeV and
      $\tan\beta=10$. This is related to the DM abundance, so that our DM scenario is
      testable at the CERN Large Hadron Collider (LHC) 
      and the International Linear Collider (ILC) by searching the missing decay of $h$.
      Furthermore, $\eta$  is potentially detectable by direct DM searches~\cite{xmass},    
      because $\eta$ can scatter with nuclei via the scalar exchange~\cite{john}. 

      Because of Type-X Yukawa interaction~\cite{barger,type-X,logan}, $H$ (or $A$) can predominantly decay
      into $\tau^+\tau^-$ instead of $b\bar b$ for $\tan\beta\gsim 2$; 
    $B(H (A) \to\tau^+\tau^-) \simeq 100$ \% and $B(H (A) \to\mu^+\mu^-)  \simeq 
    0.3$ \% for $m_A^{}=m_H^{}=130$ GeV, $\sin(\beta-\alpha)=1$ and
    $\tan\beta=10$.
      At the LHC  (30 fb$^{-1}$),
      the model can be distinguished from the MSSM Higgs sector
      by using $gg\to A (H) \to \ell^+\ell^-$ and
      $pp\to b \bar b A \to b\bar b \ell^+\ell^-$ except for the intermediate
      region of $\tan\beta$, where
       $\ell$ represents $\mu$ and $\tau$~\cite{type-X}.  
      In addition, our scenario with light $H^\pm$ and $H$ (or $A$) can be directly
      tested at the LHC (300 fb$^{-1}$) via $pp\to W^\ast \to H H^\pm$ and
      $A H^\pm$~\cite{wah}, and also $pp \to HA$. The process $e^+e^- \to HA$ at the ILC
      can also be used. Their signals are four lepton states
       $\ell^-\ell^+\tau^\pm\nu$ and $\ell^-\ell^+\tau^+\tau^-$~\cite{type-X,logan}.

For successful baryogenesis, $S^\pm$ has to have the non-decoupling property that affects
      the $hhh$ coupling~\cite{ewbg-thdm2}.  The $hhh$ coupling should deviate from the SM value by
      more than about 20~\% (see Fig.~\ref{etaOmega}), which would be tested 
      at the ILC~\cite{hhh-measurement} and its $\gamma\gamma$ option~\cite{gamgam}.
$S^\pm$ can be produced in pair at the LHC and
    the ILC, and decay into $\tau^\pm \nu \eta$. 
    The signal would be a hard hadron pair with a large missing energy~\cite{hagiwara}.
    In addition, the Majorana nature in the sub-diagram in Fig.~\ref{diag-numass}
    can be directly tested by the process $e^-e^- \to S^-S^-$ at the ILC $e^-e^-$ option 
    due to $h_e^\alpha \sim {\cal O}(1)$~\cite{aks}.
%


Finally, we comment on the case with the CP violating phases.
Our model includes the THDM, so that the same discussion can be applied
in evaluation of  baryon number at the EWPT~\cite{ewbg-thdm}.  
The mass spectrum  would be changed to some extent, but most of the features
discussed above should be conserved with a little modification. 

\section{Summary}

In this talk, we have discussed the model with the extended Higgs sector and
TeV-scale RH neutrinos, which would explain neutrino mass and mixing,
DM and baryon asymmetry by the TeV scale physics. It gives
specific predictions on the collider phenomenology. In particular, 
the predictions on the Higgs physics are completely different from those in
the MSSM, so that the model can be distinguished at the LHC and also at
the ILC.

%


\section*{References}
\begin{thebibliography}{99}

\bibitem{aks}
        M.~Aoki, S.~Kanemura and O.~Seto, Phys. Rev. Lett. {\bf 102},
        051805 (2009);
        M.~Aoki, S.~Kanemura, O.~Seto, arXiv:0904.3829 [hep-ph].
        
\bibitem{ewbg-thdm}

  J.~M.~Cline, K.~Kainulainen and A.~P.~Vischer,
  Phys.\ Rev.\  D {\bf 54}, 2451 (1996);
  L.~Fromme, S.~J.~Huber and M.~Seniuch,
  JHEP {\bf 0611}, 038 (2006).

\bibitem{zee}
  A.~Zee,
  Phys.\ Lett.\  B {\bf 93}, 389 (1980)
  [Erratum-ibid.\  B {\bf 95}, 461 (1980)];
  A.~Zee,
  Phys.\ Lett.\  B {\bf 161}, 141 (1985).

\bibitem{knt}
  L.~M.~Krauss, S.~Nasri and M.~Trodden,
  Phys.\ Rev.\  D {\bf 67}, 085002 (2003).
 \bibitem{kingman-seto}
  K.~Cheung and O.~Seto,
  Phys.\ Rev.\  D {\bf 69}, 113009 (2004).
 \bibitem{ma}
  E.~Ma,
  Phys.\ Rev.\  D {\bf 73}, 077301 (2006);
  J.~Kubo, E.~Ma and D.~Suematsu,
  Phys.\ Lett.\  B {\bf 642}, 18 (2006);
  T.~Hambye, et al., 
  Phys.\ Rev.\  D {\bf 75}, 095003 (2007); 
   K.~S.~Babu and E.~Ma,
   arXiv:0708.3790 [hep-ph]; 
   N.~Sahu and U.~Sarkar,
   arXiv:0804.2072 [hep-ph].
\bibitem{barger}
  V.~D.~Barger, J.~L.~Hewett and R.~J.~N.~Phillips,
  Phys.\ Rev.\  D {\bf 41}, 3421 (1990).
\bibitem{type-X}
  M.~Aoki, S.~Kanemura, K.~Tsumura, and K.~Yagyu, arXiv:0902.4665 [hep-ph].

\bibitem{logan}
  S.~Su and B.~Thomas,
  arXiv:0903.0667 [hep-ph]; 
        %
  H.~E.~Logan and D.~MacLennan,
  arXiv:0903.2246 [hep-ph].

  

\bibitem{lfv-data}

   M.~L.~Brooks {\it et al.}  [MEGA Collaboration],
   Phys.\ Rev.\ Lett.\  {\bf 83} (1999) 1521

  
 \bibitem{lep-data}

 T.~Schwetz, M.~Tortola and J.~W.~F.~Valle,
 New J.\ Phys.\  {\bf 10} (2008) 113011.

  
\bibitem{bsgamma}

 E.~Barberio {\it et al.}  [Heavy Flavor Averaging Group],
 arXiv:0808.1297 [hep-ex].

\bibitem{sph-cond}   
  G.~D.~Moore,
  Phys.\ Lett.\  B {\bf 439}, 357 (1998);
  Phys.\ Rev.\  D {\bf 59}, 014503 (1998).
\bibitem{xmass}
 Y.~D.~Kim,
 Phys.\ Atom.\ Nucl.\ {\bf 69}, 1970 (2006); 
D.~S.~Akerib,  et al., 
        Phys.\ Rev.\ Lett.\ {\bf 96}, 011302 (2006).
\bibitem{john}
J.~McDonald, Phys.\ Rev.\  D {\bf 50}, 3637 (1994); 
for a recent study, see {\it e.g.}, 
H.~Sung Cheon, S.~K.~Kang and C.~S.~Kim, 
  J. Cosmol. Astropart. Phys. 05 (2008) 004.

\bibitem{wah}
  S.~Kanemura and C.~P.~Yuan,
  Phys.\ Lett.\  B {\bf 530}, 188 (2002);
  Q.~H.~Cao, S.~Kanemura and C.~P.~Yuan,
  Phys.\ Rev.\  D {\bf 69}, 075008 (2004);
  A.~Belyaev, {\it et al.}, 
  Phys.\ Rev.\ Lett.\  {\bf 100}, 061801 (2008)
  [arXiv:hep-ph/0609079].

\bibitem{ewbg-thdm2}
  S.~Kanemura, Y.~Okada and E.~Senaha,
  Phys.\ Lett.\  B {\bf 606}, 361 (2005).

  
 \bibitem{hhh-measurement}
  M.~Battaglia, {\it et al.},
  arXiv:hep-ph/0111276;  
Y.~Yasui, et al., arXiv:hep-ph/0211047.

\bibitem{gamgam}
  
  R.~Belusevic and G.~Jikia,
  Phys.\ Rev.\  D {\bf 70}, 073017 (2004); 
%
  E.~Asakawa,  {\it et al.},
  arXiv:0902.2458 [hep-ph];
%
  T.~Takahashi {\it et al.},
  arXiv:0902.3377 [hep-ex].
  
\bibitem{hagiwara}
  B.~K.~Bullock, K.~Hagiwara and A.~D.~Martin,
  Phys.\ Rev.\ Lett.\  {\bf 67}, 3055 (1991).

\end{thebibliography}
\end{document}